\documentclass[epj]{webofc}
\usepackage[utf8]{inputenc}
\usepackage[varg]{txfonts}   
\usepackage{booktabs}
\usepackage{xcolor}
\definecolor{darkred}{rgb}{0.4,0.0,0.0}
\definecolor{darkgreen}{rgb}{0.0,0.4,0.0}
\definecolor{darkblue}{rgb}{0.0,0.0,0.4}
\usepackage[bookmarks,linktocpage,colorlinks,
    linkcolor = darkred,
    urlcolor  = darkblue,
    citecolor = darkgreen]{hyperref}
\usepackage{graphicx,subcaption,wrapfig}
\usepackage{bbold}

\usepackage{pgfplots}
\usepackage{cleveref}
  \usetikzlibrary{plotmarks}
  \usetikzlibrary{arrows.meta}
  \usetikzlibrary{external}
  \usepgfplotslibrary{patchplots}
  \usepackage{grffile}
\usepackage{pgfgantt}
\usepackage{pdflscape}
\pgfplotsset{compat=newest} 
\pgfplotsset{plot coordinates/math parser=false}
\pgfplotsset{grid style={dotted}}
\pgfplotsset{scaled x ticks=false}
\pgfplotsset{scaled y ticks=false}
\newlength\figureheight 
\newlength\figurewidth 
\wocname{EPJ Web of Conferences}
\woctitle{Lattice2017}
\newcommand{\SU}[1]{\text{SU}(#1)}
\newcommand{\cl}{\text{clover}}
\newcommand{\pl}{\text{plaq}}
\newcommand{\plat}{\text{plat}}
\newcommand{\tr}{\mathop{\text{tr}}}
\renewcommand{\Re}{\mathop{\text{Re}}}
\newcommand{\Exp}[1]{\left\langle #1\right\rangle}
\newcommand{\ord}[1]{\mathrm{O}\left(#1\right)}

\newcommand{\aqq}{\alpha_{\rm qq}}
\newcommand{\gqqt}{\overline{g}^2_{\rm qq}}
\newcommand{\gqq}{\overline{g}_{\rm qq}}
\newcommand{\qq}{{\rm qq}}
\newcommand{\fm}{\mathrm{fm}}
 \newcommand{\MSbar}{\overline{\rm{MS}}}
 \newcommand{\LMS}{\Lambda_{\MSbar}}
 \newcommand{\Lqq}{\Lambda_{\qq}}
 \newcommand{\slope}{\rho}
\begin{document}
%
\selectlanguage{english}
\title{%
\SU{3} Yang Mills theory at small distances and fine lattices
}
\author{%
\firstname{Nikolai} \lastname{Husung}\inst{1,2} \and
\firstname{Mateusz} \lastname{Koren}\inst{1} \and
\firstname{Philipp}  \lastname{Krah}\inst{1,2} \and
\firstname{Rainer} \lastname{Sommer}\inst{1,2}
\fnsep\thanks{Speaker, \email{rainer.sommer@desy.de}}
}
\institute{%
John von Neumann Institute for Computing (NIC), DESY, Platanenallee~6, 15738 Zeuthen, Germany
\and
Institut~f\"ur~Physik, Humboldt-Universit\"at~zu~Berlin, Newtonstr.~15, 12489~Berlin, Germany
}
\abstract{%
  We investigate the SU(3) Yang Mills theory at small gradient flow time and at short distances. Lattice spacings down to $a=0.015\,$fm are simulated with open boundary conditions to allow topology to flow in and out. We study the behaviour of the action density $E(t)$ close to the boundaries, the feasibility of the small flow-time expansion and the extraction of the $\Lambda$-parameter from the static force at small distances. For the latter, significant deviations from the 4-loop perturbative $\beta$-function are visible at $\alpha\approx 0.2\,$. We still can extrapolate to extract $r_0\Lambda$.
}
\hfill DESY 17-176
\maketitle
\section{Introduction}\label{intro}

To simulate \SU{3} Yang Mills theory on fine lattices, close to the continuum limit with lattice spacings down to $a=0.015\,$fm one has to avoid the freezing of the topological charge \cite{hmc:slowingdown,Luscher_TopoWflowHMC}. Here this is achieved by introducing open boundary conditions in time \cite{Schaefer_LQCDwoTopologyBarriers}. These boundary conditions allow the topological charge to flow in and out of the lattice, but break translation invariance explicitly. 

The resulting boundary effects have been seen to extend over a large region \cite{Bruno:2014ova}, significantly reducing the volume which is accessible for the determination of vacuum expectation values. The latter is referred to as plateau. 
Deviations near the time boundaries from the plateau values 
are caused both by boundary conditions affecting the continuum theory
and by discretisation effects. Only the latter can  be reduced by
boundary improvement terms.

To find out whether the plateau region can be enlarged at finite lattice spacing by adjusting boundary counterterms, we here consider 
the action density at positive flow time $t$
\begin{equation}
E(t,x)=-\frac{1}{2}\tr\left(G_{\mu\nu}(t,x)G_{\mu\nu}(t,x)\right)
\end{equation}
and evaluate it close to the time boundary. Additionally the deviation of the action density near -- but not too close to -- the time boundary is used as a testing ground for the small flow-time
expansion \cite{LuscherWeisz_PertFlow}. 
This deviation yields specific matrix elements of the energy-momentum tensor analogously to what is done at finite temperature \cite{Ejiri_LatentHeat}.
Restricting attention to the plateau region, 
the accessible fine lattices can be used to go down to small distances, probing the perturbative regime.
With the coupling in the $\qq$-scheme 
\begin{equation}
	\gqqt(r)=3\pi r^2 F(r)\,,
\end{equation}
defined in terms of the static force, it is possible to compare 
the non-perturbative running with the perturbative expansion. 
The Renormalization Group $\beta$-function in the $\qq$-scheme
\begin{equation}
	\beta^\mathrm{qq}(\gqq) \equiv -r\frac{\mathrm{d}}{\mathrm{d} r} \gqq(r)\,,
\label{eq:RGE}
\end{equation}
is perturbatively known up to 4-loop order \cite{Peter:1997me,Schroder:1998vy,Anzai:2009tm,Smirnov:2009fh,Brambilla:1999xf},
\begin{equation}
  		\beta^{\rm qq}(\gqq)=-\gqq^3\left[b_0+b_1\gqq^2 +b_2\gqq^4+\left(b_3+b_{3,\rm{IR}}\log\left(C_\mathrm{A}\frac{\gqqt}{8\pi}\right)\right)\gqq^6 \right]+\ord{\gqq^{11}}\,.
  \end{equation} 
Starting at 4 loops the coupling is not infrared safe.
The resummation of the infrared divergence leads to the $\log(g^2)$ term
\cite{Brambilla:1999xf}.   
From the integration of \cref{eq:RGE} with the 
perturbative $\beta^\mathrm{qq}$ we will extract the $\Lambda$-parameter and test perturbation theory from the 
condition that $\Lambda$ is Renormalization Group Invariant.

\section{Simulation details}\label{sec:simulation}
We use the Wilson action \cite{Wilson_ConfinementQuarks} with open boundary conditions in time, exactly as described in  \cite{Schaefer_LQCDwoTopologyBarriers}. In particular the boundary 
O$(a)$ counterterms are implemented as written there and
their coefficients are set to the tree-level values.
The gauge field configurations are generated using local updates, namely one pseudo heat-bath sweep \cite{Cabibbo_PseudoHeatBath} followed by multiple over-relaxation sweeps \cite{Creutz:1987,Brown:1987,Adler:1988} acting on \SU{2} subgroups.

The database of our analysis consists of $N_\text{wl}$ Wilson loop measurements and $N_\text{flow}$ Wilson flow measurements listed in \cref{tabl:database}. It is presently being enlarged.
\begin{table}[htp]
\centering
\begin{tabular}{c c c c c c}
\toprule
$\beta$		& $a\,$[fm]		&$L/a$  & $r/r_0$ &$N_\text{wl}$	&$N_\text{flow}$\\
\midrule
6.0662		&0.0834(4)		&24     &$[0.42, 1.92]$ &121	&511\\
6.2556		&0.0624(4)		&32    &$[0.31, 1.44]$ &101	&361\\
6.3406		&{\color{white!30!black}\textit{0.0555(2)}}&36       	&--				&--	&341\\
6.5619		&0.0411(2)		&48     &$[0.21, 1.04]$	&301	&165\\
6.7859		&0.0312(2)		&64    &$[0.16, 1.22]$	&64	&49\\
6.9606		&{\color{white!30!black}\textit{0.0250(3)}}&80			&--				&--	&76\\
7.1146		&0.0206(2)		&96     &$[0.10, 0.64]$	&64	&--\\
7.3600		&0.0152(2)		&128     &$[0.07, 0.48]$	&58	&--\\
\bottomrule
\end{tabular}
\caption{List of the ensembles used for our analysis. Here $\beta$ is the inverse coupling, $a$ is the lattice spacing, defined by $r_0=0.5\,\fm$  and $r/r_0$ the distance regime of the computed quantities. The statistics for the Wilson loop and flow measurement are given in the last two columns. The lattice spacings in italic font were computed from the parametrization of $r_0/a$ given in \cite{Necco_heavyQuarkFromShortToIntermediate}.}
\label{tabl:database}
\end{table}
To keep track of autocorrelations, we measured the topological charge $Q(t)$. Within large errors, the autocorrelation function of $Q^2(t_0)$ is in agreement with scaling in the variable
$t_\mathrm{MC} \times a^2$, with the Monte Carlo time, $t_\mathrm{MC}$, defined
in units of sweeps.

\section{Open boundaries and small flow time}\label{sec:actionDensity}
The Wilson flow \cite{Luscher_PropUsesWilsonFlowLQCD} is employed to determine the scale $t_0$ and to obtain the deviation of the action density near the boundary from the plateau value $t_0^2\Exp{E(t,x)-E(t,x^\plat)}$, where $t=t_0$ as well as small flow times $t/t_0\in[0.06,0.17]$. The action density is considered in both the clover and plaquette discretisation
\begin{equation}\label{eq:actionDensity}
E_\cl(t,x)=-\frac{1}{2}\tr\Big(G_{\mu\nu}^\text{lat}(t,x)G_{\mu\nu}^\text{lat}(t,x)\Big)\,\,,\,\,E_\pl(t,x)=\frac{1}{a^4}\sum\limits_{\mu,\nu}\Re\tr\Big(\mathbb{1}-V_{\mu\nu}(t,x)\Big)\,,
\end{equation}
where $V_{\mu\nu}$ is the plaquette and $G_{\mu\nu}^\text{lat}$ is the clover definition of the field strength tensor \cite{Sheikholeslami:1985ij} computed at flow time $t$.
\subsection{The action density near the boundary}\label{sec:boundCT}
\begin{figure}[hb] 
\centering
\footnotesize
\setlength\figureheight{4.5cm}
\setlength\figurewidth{0.4\textwidth}
\begin{subfigure}{0.48\textwidth}
\input{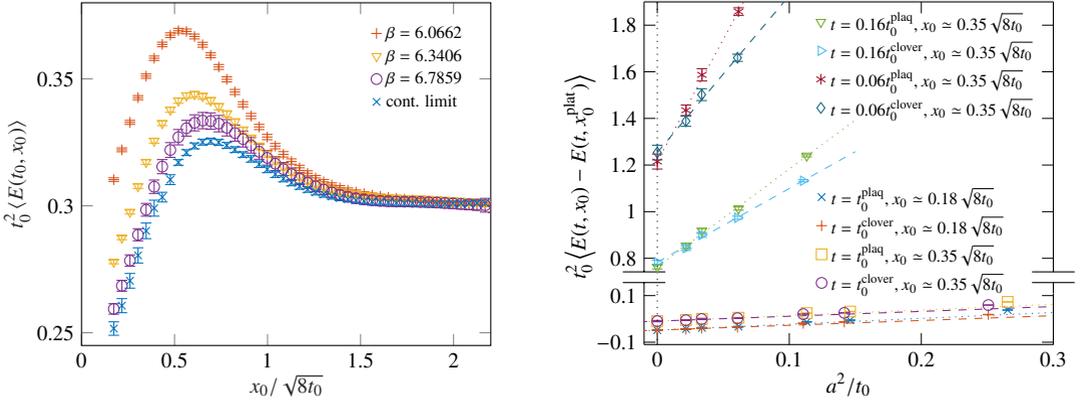}
\caption{Boundary peak for different lattice spacings at ${t=t_0}$.}\label{fig:bndPeak}
\end{subfigure}\hfill
\begin{subfigure}{0.48\textwidth}
\vspace{-0.225cm}\hspace{-2.5cm}
%
\definecolor{mycolor1}{rgb}{0.00000,0.44700,0.74100}%
\definecolor{mycolor2}{rgb}{0.85000,0.32500,0.09800}%
\definecolor{mycolor3}{rgb}{0.92900,0.69400,0.12500}%
\definecolor{mycolor4}{rgb}{0.49400,0.18400,0.55600}%
\definecolor{mycolor5}{rgb}{0.46600,0.67400,0.18800}%
\definecolor{mycolor6}{rgb}{0.30100,0.74500,0.93300}%
\definecolor{mycolor7}{rgb}{0.63500,0.07800,0.18400}%

\usetikzlibrary{pgfplots.groupplots}
\begin{tikzpicture}
\begin{groupplot}[
group style={
    group name=my fancy plots,
    group size=1 by 2,
    xticklabels at=edge bottom,
    vertical sep=0pt
},
xlabel shift = -3 pt,
ylabel shift = -3 pt,
width=0.951\figurewidth,
height=\figureheight,
at={(0\figurewidth,0\figureheight)},
scale only axis,
separate axis lines,
every outer x axis line/.append style={black},
every x tick label/.append style={font=\color{black},inner sep=0.333em,outer sep=0.2\pgflinewidth},
every y tick label/.append style={font=\color{black},inner sep=0.333em,outer sep=0.2\pgflinewidth},
every x tick/.append style={black},
xmin=-0.01,
xmax=0.3,
ytick ={-0.1,-0.05,0,0.05,0.1},
every outer y axis line/.append style={black},
every y tick label/.append style={font=\color{black}},
every y tick/.append style={black},
axis background/.style={fill=white}
]
\nextgroupplot[
ymin=0.65,
ymax=1.9,
height=0.850340136\figureheight,
axis x line*=top,
axis y discontinuity=parallel,
ylabel={$\displaystyle t_0^2\left\langle E(t,x_0)-E(t,x_0^\mathrm{plat})\right\rangle$},
y label style={at={(-0.1,0.4)}},
ytick={0.8,1,1.2,1.4,1.6,1.8},
legend style={at={(0.4,1)}, anchor=north west, nodes={scale=0.8, transform shape}, legend cell align=left, align=left, draw=none, fill=none},
scaled ticks=false,tick label style={/pgf/number format/fixed}
]
\addplot [only marks,color=mycolor5, draw=none, mark=triangle, mark options={solid, rotate=180, mycolor5}]
 plot [error bars/.cd, y dir = both, y explicit]
 table[row sep=crcr, y error plus index=2, y error minus index=3]{%
0	0.762595910046676	0.00863498099732553	0.00863498099732553\\
0.113170331807838	1.23959511444376	0.00477854799952649	0.00477854799952649\\
0.061544925270555	1.01343507500047	0.00689772364622269	0.00689772364622269\\
0.033843123634618	0.917480847558795	0.010482415641639	0.010482415641639\\
0.0217939374240917	0.853506947161118	0.0104275960860001	0.0104275960860001\\
};
\addlegendentry{$\displaystyle t=0.16t_0^\mathrm{plaq},x_0\simeq 0.35\sqrt{8t_0}$}

\addplot [color=mycolor5, dotted, forget plot, domain=0:0.15]
  table[row sep=crcr]{%
0	0.762595910046676\\
0.15 1.393111012371095\\
};
\addplot [only marks,color=mycolor6, draw=none, mark=triangle, mark options={solid, rotate=270, mycolor6}]
 plot [error bars/.cd, y dir = both, y explicit]
 table[row sep=crcr, y error plus index=2, y error minus index=3]{%
0	0.783113629109	0.00863008093644438	0.00863008093644438\\
0.11053988978432	1.13259304704081	0.00464054653822825	0.00464054653822825\\
0.0607735480425526	0.974062528803852	0.00682853726622043	0.00682853726622043\\
0.033608831718394	0.900278617678077	0.010407619259667	0.010407619259667\\
0.0216963242656688	0.8439322056019	0.0103617669542862	0.0103617669542862\\
};
\addlegendentry{$\displaystyle t=0.16t_0^\mathrm{clover},x_0\simeq 0.35\sqrt{8t_0}$}

\addplot [color=mycolor6, dashed, forget plot]
  table[row sep=crcr]{%
0	0.783113629108919\\
0.15 1.257408127579358\\
};
\addplot [only marks,color=mycolor7, draw=none, mark=asterisk, mark options={solid, mycolor7}]
 plot [error bars/.cd, y dir = both, y explicit]
 table[row sep=crcr, y error plus index=2, y error minus index=3]{%
0	1.21507135953315	0.0321991558751749	0.0321991558751749\\
0.061544925270555	1.85816964351264	0.0165715664336063	0.0165715664336063\\
0.033843123634618	1.58694186218374	0.026362230626748	0.026362230626748\\
0.0217939374240917	1.43484119981515	0.0225013993343321	0.0225013993343321\\
};
\addlegendentry{$\displaystyle t=0.06t_0^\mathrm{plaq},x_0\simeq 0.35\sqrt{8t_0}$}

\addplot [color=mycolor7, dotted, forget plot]
  table[row sep=crcr]{%
0	1.21507135953315\\
0.35	4.88396226198185\\
};
\addplot [only marks,color=teal!80!violet, draw=none, mark=diamond, mark options={solid, teal!80!violet}]
 plot [error bars/.cd, y dir = both, y explicit]
 table[row sep=crcr, y error plus index=2, y error minus index=3]{%
0	1.25361227190	0.0313873165978352	0.0313873165978352\\
0.0607735480425526	1.66022623618495	0.0153842523744103	0.0153842523744103\\
0.033608831718394	1.50188639224836	0.0254072602113989	0.0254072602113989\\
0.0216963242656688	1.38815354421076	0.0220093443290112	0.0220093443290112\\
};
\addlegendentry{$\displaystyle t=0.06t_0^\mathrm{clover},x_0\simeq 0.35\sqrt{8t_0}$}

\addplot [color=teal!80!violet, dashed, forget plot]
  table[row sep=crcr]{%
0	1.2536122771902\\
0.35	3.60952752348576\\
};
\addplot [color=black, dotted, forget plot]
  table[row sep=crcr]{%
0	-1\\
0	5\\
};
\nextgroupplot[
height=0.149659864\figureheight,
ymin=-0.11,
ymax=0.11,
ytick={-0.1,0.1},
axis x line*=bottom,
xlabel={$\displaystyle a^2/t_0$},
legend style={at={(0.40,0.81)}, anchor=south west, nodes={scale=0.8, transform shape}, legend cell align=left, align=left, draw=none, fill=none},
scaled ticks=false,tick label style={/pgf/number format/fixed}
]
\addplot [only marks,color=mycolor1, draw=none, mark=x, mark options={solid, mycolor1}]
 plot [error bars/.cd, y dir = both, y explicit]
 table[row sep=crcr, y error plus index=2, y error minus index=3]{%
0	-0.048148970646766	0.00223464248678052	0.00223464248678052\\
0.265395125722936	0.0379810428014566	0.000752678768265698	0.000752678768265698\\
0.146056532937281	-0.00520289832777132	0.000727789856693091	0.000727789856693091\\
0.113170331807838	-0.013020088628966	0.00078545890237095	0.00078545890237095\\
0.061544925270555	-0.0327475352057591	0.00107185644875724	0.00107185644875724\\
0.033843123634618	-0.0382002399908692	0.00212207537844102	0.00212207537844102\\
0.0217939374240917	-0.0431684045003251	0.00152707112793265	0.00152707112793265\\
};
\addlegendentry{$\displaystyle t=t_0^\mathrm{plaq},x_0\simeq 0.18\sqrt{8t_0}$}

\addplot [color=mycolor1, dotted, forget plot]
  table[row sep=crcr]{%
-0.05	-0.0607504876191663\\
0.3	0.0274601311876363\\
};
\addplot [only marks,color=mycolor2, draw=none, mark=+, mark options={solid, mycolor2}]
 plot [error bars/.cd, y dir = both, y explicit]
 table[row sep=crcr, y error plus index=2, y error minus index=3]{%
0	-0.0480848175214183	0.00224656883337234	0.00224656883337234\\
0.2507694850678	0.0186504847636227	0.000728857848958931	0.000728857848958931\\
0.141661236097228	-0.0133079860302241	0.000730784594696267	0.000730784594696267\\
0.11053988978432	-0.0189282230862688	0.000777911284543005	0.000777911284543005\\
0.0607735480425526	-0.0356243436874734	0.0010740967367298	0.0010740967367298\\
0.033608831718394	-0.0397107678549798	0.00211982387041308	0.00211982387041308\\
0.0216963242656688	-0.044108930375912	0.00152618100382527	0.00152618100382527\\
};
\addlegendentry{$\displaystyle t=t_0^\mathrm{clover},x_0\simeq0.18\sqrt{8t_0}$}

\addplot [color=mycolor2, dashed, forget plot]
  table[row sep=crcr]{%
-0.05	-0.0584279191763546\\
0.3	0.0139737924081994\\
};
\addplot [only marks,color=mycolor3, draw=none, mark=square, mark options={solid, mycolor3}]
 plot [error bars/.cd, y dir = both, y explicit]
 table[row sep=crcr, y error plus index=2, y error minus index=3]{%
0	-0.00947142037342343	0.00260798363870268	0.00260798363870268\\
0.265395125722936	0.0729203120662234	0.000901604473892643	0.000901604473892643\\
0.146056532937281	0.0315627602146984	0.00095087521112184	0.00095087521112184\\
0.113170331807838	0.0248406734082477	0.000895863479139987	0.000895863479139987\\
0.061544925270555	0.00513956765927089	0.00128493894998136	0.00128493894998136\\
0.033843123634618	0.000517840577921625	0.00260387975639279	0.00260387975639279\\
0.0217939374240917	-0.00484593344937314	0.00175924953044838	0.00175924953044838\\
};
\addlegendentry{$\displaystyle t=t_0^\mathrm{plaq},x_0\simeq0.35\sqrt{8t_0}$}

\addplot [color=mycolor3, dotted, forget plot]
  table[row sep=crcr]{%
-0.05	-0.0214542509001537\\
0.3	0.0624255627869581\\
};
\addplot [only marks,color=mycolor4, draw=none, mark=o, mark options={solid, mycolor4}]
 plot [error bars/.cd, y dir = both, y explicit]
 table[row sep=crcr, y error plus index=2, y error minus index=3]{%
0	-0.0094542199273258	0.00262311479675587	0.00262311479675587\\
0.2507694850678	0.0594820764944358	0.000883379295100058	0.000883379295100058\\
0.141661236097228	0.0258391642962999	0.000952290148001108	0.000952290148001108\\
0.11053988978432	0.0206368480527569	0.000892043256635318	0.000892043256635318\\
0.0607735480425526	0.00306535946629634	0.00128826557718626	0.00128826557718626\\
0.033608831718394	-0.000577744563383699	0.00260466626317144	0.00260466626317144\\
0.0216963242656688	-0.00552924788281511	0.00175879165628192	0.00175879165628192\\
};
\addlegendentry{$\displaystyle t=t_0^\mathrm{clover},x_0\simeq0.35\sqrt{8t_0}$}

\addplot [color=mycolor4, dashed, forget plot]
  table[row sep=crcr]{%
-0.05	-0.0198694818732103\\
0.3	0.0530373517479811\\
};

\addplot [color=black, dotted, forget plot]
  table[row sep=crcr]{%
0	-1\\
0	5\\
};
\end{groupplot}
\end{tikzpicture}%
\vspace{-3.6cm}
\caption{Examples of continuum extrapolations at flow times ${t\in\lbrace 0.06,0.16,1\rbrace t_0}$.}\label{fig:contExtrBnd}
\end{subfigure}
\caption{Action density near the time boundary for different flow times. On the right hand side, the continuum limit is taken
by an extrapolation linearly in $a^2$, omitting
O$(a)$ terms which in principle are present
due to imprecisions of the boundary
counterterms. The data supports their smallness.}
\label{fig:bndEffects}
\end{figure}

The action density at $t=t_0$ is first used as a test quantity to determine whether the extent of the plateau region increases for smaller lattice spacings and to find the overall shape of its $x_0$-dependence. 
Three examples at finite lattice spacings and the obtained continuum limit are shown in figure \ref{fig:bndPeak}. The deviation near the boundary is 
significantly affected by discretisation effects as the height of
the peak shrinks towards the continuum but the width of the  boundary region is roughly constant at around $\sqrt{20t_0}$. Hence there is no way to increase the plateau region significantly by more precise coefficients
of the used boundary counterterms. Additionally the discretisation effects are quite small at $t=t_0$, see also figure \ref{fig:contExtrBnd}.
\subsection{Small flow-time expansion}\label{sec:boundSFT}
Since $\Exp{E(t,x)}$ in the boundary region has a smooth continuum limit, it can be used as a test case for the small flow-time expansion \cite{LuscherWeisz_PertFlow,Luscher_FutureApplicationsFlow}. More precisely the trace of the energy-momentum tensor $T_{\mu\mu}(x)$ can be extracted from the expansion of the action density \cite{LuscherWeisz_PertFlow,Suzuki_EnergyMomentumTensor,Ejiri_LatentHeat}
\begin{equation}
E(t,x)=c_\mathbf{1}(t)+c_E(t)T_{\mu\mu}(x)+\ord{t}\,,
\end{equation}
with coefficients \cite{Suzuki_EnergyMomentumTensor} 
\begin{equation}
c_\mathbf{1}(t)=\left\langle E(t,x_0^\text{plat})\right\rangle\,,\,c_E(t)=\frac{1}{2b_0}\Big(1+2b_0\bar{s} \bar{g}^2 + \ord{\bar{g}^4}\Big)\,.
\end{equation}
The definition of $c_\mathbf{1}(t)$ sets the vacuum expectation 
value of the trace to zero. The deviation of the action density 
at small flow time from its vacuum expectation value then yields
\begin{equation}
\Exp{T_{\mu\mu}(x_0)}=\lim_{t\rightarrow 0}\frac{1}{c_E(t)}
\lim_{a\to0}\Exp{E(t,x_0)-E(t,x_0^\plat)}\,.
\end{equation}
The continuum limit and zero flow-time limit have to be performed in the correct order. Note that $\Exp{T_{\mu\mu}(x_0)}$ may be expressed as
a Hilbert space off-diagonal matrix element between 
a $x_0$-dependent boundary state and the vacuum. This is qualitatively similar to the finite temperature application.
The lattice data entering the extrapolations are restricted in flow time by \cite{Luscher_FutureApplicationsFlow}
\begin{equation}\label{eq:restrictionSFT}
a\ll \sqrt{8t}\ll \text{relevant low energy scales}\,.
\end{equation}
\begin{figure}[t]
\centering
\footnotesize
\setlength\figureheight{4.5cm}
\setlength\figurewidth{0.4\textwidth}
\sidecaption
\input{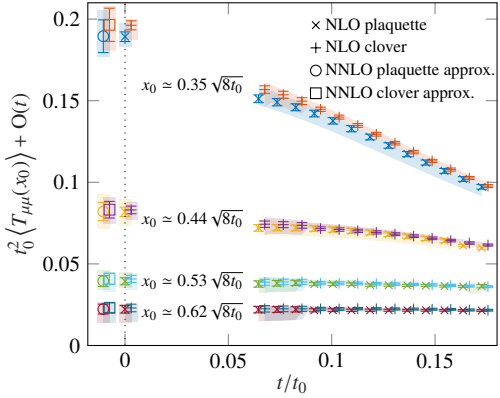}
\caption{Trace of the energy-momentum tensor plus small flow-time corrections. Both discretizations, i.e., plaquette and clover are included, where the clover definition is overall shifted by $0.003$ in $t/t_0$ for readability. The coefficient $c_E(t)$ is known up to next to leading order. Uncertainties due to NNLO effects 
are estimated with \cref{eq:NNLOapprox}. Their
effect is included in the error at zero flow time (shifted by $-0.01$ in $t/t_0$). The shaded areas mark the results of continuum extrapolations omitting one more lattice spacing.}
\label{fig:sftLimit}
\end{figure}
As it turns out, discretisation effects are large for flow times $t\simeq 0.06t_0$ close to the boundary, where the matrix element is large. Hence the analysis has been restricted to $x_0\gtrsim 0.35\sqrt{8t_0}$ where a continuum extrapolation seems feasible, see figure \ref{fig:contExtrBnd}. Depending on the chosen flow time, different numbers of lattice spacings $a\in[0.025,0.056]\,\text{fm}$ can be included into linear continuum extrapolations due to equation \eqref{eq:restrictionSFT} and discretisation effects of higher order in the lattice spacing. Selected continuum extrapolated values at positive flow time are shown in \cref{fig:sftLimit}. Also the zero flow-time limit was performed. \Cref{fig:sftLimit} has been updated 
from the one shown at the conference, where correlations 
in $t$ were not yet included. As $c_E(t)$ is only known up to next to leading order (NLO) the NNLO effects are roughly estimated via
\begin{equation}\label{eq:NNLOapprox}
\Delta c_E(t)=\pm\frac{\bar{g}^4}{2b_0(4\pi)^2}=\pm\text{LO}\times \alpha^2\,.
\end{equation}
The results show no significant impact of $\Delta c_E(t)$ on the zero flow-time limit at the available accuracy of ${10-20\,\%}$.

\newpage
\section{Step scaling for short distances and large volume} 
\label{sec:step_scaling_for_short_distances_and_large_volume}
For six different lattice spacings $a$, Wilson loops have been measured with total statistic of $N_\text{wl}$, listed in \cref{tabl:database}. 
The coupling $\gqqt(r,a)$ at finite lattice spacing $a$ was derived from Wilson loops applying the analysis described in \cite{DetStat}
with only one difference: the parallel transporters in time are
the dynamical gauge fields (no smearing) and statistical errors
are reduced by the multi-hit technique~\cite{Parisi}. 

To extrapolate the coupling to its continuum value we used two different strategies. In the regime of intermediate distances the scale was set at $r_0$ \cite{Sommer_parameter} and the coupling $\gqqt(r,a)$ was extrapolated to the continuum at $r/r_0=0.3,0.4,\dots,1.1$. 
In the short distance regime ($r\le0.45r_0$) the continuum extrapolation of the coupling $\gqqt(r)$ was performed using step scaling.

Originally used to bridge large scale differences 
in finite volume couplings \cite{luscher1991numerical}, we use it here to extrapolate from large to small distances, in large volume. 

\begin{wrapfigure}[30]{r}{0.45\textwidth}
	\setlength\figureheight{0.5\linewidth} 
	\setlength\figurewidth{0.7\linewidth}
%
\definecolor{mycolor1}{rgb}{0.00000,0.44700,0.74100}%
\begin{tikzpicture}

\begin{axis}[%
width=0.978\figurewidth,
height=\figureheight,
at={(0\figurewidth,0\figureheight)},
scale only axis,
xmin=0,
xmax=1.14,
xlabel style={font=\color{white!15!black}},
xlabel={$r/r_0$},
ymin=0,
ymax=20,
ylabel style={font=\color{white!15!black}},
ylabel={$\gqqt(r)$},
inner sep=0.333em, outer sep=0.5\pgflinewidth,
axis background/.style={fill=white},
legend style={at={(0.03,0.97)}, anchor=north west, legend cell align=left, align=left, fill=none, draw=none}
]
\addplot [color=mycolor1]
 plot [error bars/.cd, y dir = both, y explicit]
 table[row sep=crcr, y error plus index=2, y error minus index=3]{%
0.3	3.68429776966882	0.0269884591094442	0.0269884591094442\\
0.4	4.70884167292034	0.0311220180705063	0.0311220180705063\\
0.45	5.28012775310259	0.0392534214894686	0.0392534214894686\\
0.5	5.90500708013021	0.0454556834403736	0.0454556834403736\\
0.6	7.32756973324075	0.0688946229034697	0.0688946229034697\\
0.7	8.93731168200892	0.0963503909194441	0.0963503909194441\\
0.8	10.8672247438704	0.129728349531481	0.129728349531481\\
0.9	12.986404784798	0.169829620761483	0.169829620761483\\
1.1	17.7943312667764	0.319765912539353	0.319765912539353\\
1.2	21.1217346677584	0.597086722492507	0.597086722492507\\
1.3	24.8117673165059	1.26049432068488	1.26049432068488\\
};
\addlegendentry{continuum limit}

\addplot [color=red, draw=none, only marks, mark=asterisk, mark options={solid, red}]
 plot [error bars/.cd, y dir = both, y explicit]
 table[row sep=crcr, y error plus index=2, y error minus index=3]{%
0.45	5.28012775310259	0.0392534214894686	0.0392534214894686\\
};
\addlegendentry{step scaling}
\addplot [color=red, draw=none, mark=asterisk, mark options={solid, red}]
 plot [error bars/.cd, y dir = both, y explicit]
 table[row sep=crcr, y error plus index=2, y error minus index=3]{%
0.3375	4.04214152526816	0.0279337050047957	0.0279337050047957\\
0.45	5.28012775310259	0.0392534214894686	0.0392534214894686\\
};
\addplot [color=red, draw=none, mark=asterisk, mark options={solid, red}]
 plot [error bars/.cd, y dir = both, y explicit]
 table[row sep=crcr, y error plus index=2, y error minus index=3]{%
0.253125	3.29708875840115	0.0208719593498654	0.0208719593498654\\
0.3375	4.04214152526816	0.0279337050047957	0.0279337050047957\\
0.45	5.28012775310259	0.0392534214894686	0.0392534214894686\\
};
\addplot [color=red, draw=none, mark=asterisk, mark options={solid, red}]
 plot [error bars/.cd, y dir = both, y explicit]
 table[row sep=crcr, y error plus index=2, y error minus index=3]{%
0.106787109375	2.1362272592905	0.0101284281871533	0.0101284281871533\\
0.1423828125	2.41623350415678	0.0132436622951793	0.0132436622951793\\
0.18984375	2.78375180858234	0.0145980216819528	0.0145980216819528\\
0.253125	3.29708875840115	0.0208719593498654	0.0208719593498654\\
0.3375	4.04214152526816	0.0279337050047957	0.0279337050047957\\
0.45	5.28012775310259	0.0392534214894686	0.0392534214894686\\
};

\addplot [color=red, dashed, forget plot]
  table[row sep=crcr]{%
0.45	0\\
0.45	24.8117673165059\\
};
\end{axis}

\begin{axis}[%
width=1.262\figurewidth,
height=1.262\figureheight,
at={(-0.164\figurewidth,-0.167\figureheight)},
scale only axis,
xmin=0,
xmax=1,
ymin=0,
ymax=1,
axis line style={draw=none},
ticks=none,
axis x line*=bottom,
axis y line*=left
]
\node[below right, align=left, font=\color{red}, draw=white]
at (rel axis cs:0.15,0.6) {step scaling\\$r\le0.45r_0$};
\end{axis}
\end{tikzpicture}%
	\caption{Six step scaling iterations starting from $r_*=0.45r_0$ reaching down to $r\approx0.1r_0$ and the continuum limit for the large distance regime.}
  \label{fig:step_scaling_iter}
  \setlength\figureheight{0.5\linewidth} 
      \setlength\figurewidth{0.7\linewidth}
%
\begin{tikzpicture}

\begin{axis}[%
width=0.978\figurewidth,
height=\figureheight,
at={(0\figurewidth,0\figureheight)},
scale only axis,
xmin=-0.004,
xmax=0.110655805411185,
xtick={0,0.025,0.05,0.075,0.1},
xticklabels={0,0.025,0.05,0.075,0.1},
xlabel style={font=\color{white!15!black}},
xlabel={$(a/r)^2$},
ymin=0.75,
ymax=0.9,
ylabel style={font=\color{white!15!black}},
ylabel={$\gqqt(s^kr_*,a/r)/u$},
axis background/.style={fill=white},
legend style={at={(0.97,0.5)}, anchor=east, legend cell align=left, align=left, fill=none, draw=none},
scaled ticks=false,
]
\addplot [color=black, draw=none, mark=+, mark options={solid, black}, forget plot]
 plot [error bars/.cd, y dir = both, y explicit]
 table[row sep=crcr, y error plus index=2, y error minus index=3]{%
0	0.765046655208972	0.00586046941274114	0.00586046941274114\\
0.0786775433548226	0.768488386896565	0.00625510931687292	0.00625510931687292\\
0.0339185118425725	0.767494804389472	0.00478791302100817	0.00478791302100817\\
0.0190206246667431	0.766289614921403	0.00772960752320639	0.00772960752320639\\
0.00838142581815558	0.764397885602645	0.00653616046396089	0.00653616046396089\\
};
\addplot [color=black, dotted, forget plot]
  table[row sep=crcr]{%
0	0.765046654746478\\
0.0786775433548226	0.768983006647028\\
};
\addplot [color=black, draw=none, mark=+, mark options={solid, black}, forget plot]
 plot [error bars/.cd, y dir = both, y explicit]
 table[row sep=crcr, y error plus index=2, y error minus index=3]{%
0	0.817565905184932	0.00548401620080615	0.00548401620080615\\
0.0607761481795061	0.811671001208036	0.00516758477289762	0.00516758477289762\\
0.0339467941379853	0.814822409717802	0.00580954391887278	0.00580954391887278\\
0.0148689088994874	0.814602525815459	0.0052001728848758	0.0052001728848758\\
0.00804344164720959	0.81889855789297	0.00718024238474043	0.00718024238474043\\
};
\addplot [color=black, dotted, forget plot]
  table[row sep=crcr]{%
0	0.817565905075178\\
0.0607761481795061	0.811686211772619\\
};
\addplot [color=black, draw=none, mark=+, mark options={solid, black}, forget plot]
 plot [error bars/.cd, y dir = both, y explicit]
 table[row sep=crcr, y error plus index=2, y error minus index=3]{%
0	0.845693731382876	0.00456658105843861	0.00456658105843861\\
0.105655805411185	0.847313409817725	0.0045725776146936	0.0045725776146936\\
0.0597294958283814	0.841869066307128	0.00521896002514551	0.00521896002514551\\
0.0261420955349659	0.844831992434105	0.00455905643712709	0.00455905643712709\\
0.0143727328275552	0.848095502192982	0.00471969699452961	0.00471969699452961\\
};
\addplot [color=black, dotted, forget plot]
  table[row sep=crcr]{%
0	0.845693731919484\\
0.105655805411185	0.845733083053316\\
};
\addplot [color=black, draw=none, mark=+, mark options={solid, black}, forget plot]
 plot [error bars/.cd, y dir = both, y explicit]
 table[row sep=crcr, y error plus index=2, y error minus index=3]{%
0	0.867138560046901	0.00432309995188447	0.00432309995188447\\
0.104416182043728	0.872517465006013	0.00355818404849582	0.00355818404849582\\
0.0462931379316308	0.869107068169742	0.00407263279790639	0.00407263279790639\\
0.0258270057403625	0.868722803055003	0.00399770581925968	0.00399770581925968\\
};
\addplot [color=black, dotted, forget plot]
  table[row sep=crcr]{%
0	0.86713855994681\\
0.104416182043728	0.872441424886887\\
};
\addplot [only marks,color=black, draw=none, mark=+, mark options={solid, black}]
 plot [error bars/.cd, y dir = both, y explicit]
 table[row sep=crcr, y error plus index=2, y error minus index=3]{%
0	0.885057426977858	0.00415041558480335	0.00415041558480335\\
0.0831299268239188	0.8890267408201	0.00353943111823408	0.00353943111823408\\
0.0462894299855906	0.887267669212048	0.00378284238003647	0.00378284238003647\\
};
\addlegendentry{$\slope\ne0$}

\addplot [color=black, dotted, forget plot]
  table[row sep=crcr]{%
0	0.885057427118456\\
0.0831299268239188	0.889026740786837\\
};
\addplot [color=red, draw=none, mark=o, mark options={solid, red}, forget plot]
 plot [error bars/.cd, y dir = both, y explicit]
 table[row sep=crcr, y error plus index=2, y error minus index=3]{%
0.003	0.76638860859402	0.00477355215561452	0.00477355215561452\\
0.0369185118425725	0.767494804389472	0.00478791302100817	0.00478791302100817\\
0.0220206246667431	0.766289614921403	0.00772960752320639	0.00772960752320639\\
0.0113814258181556	0.764397885602645	0.00653616046396089	0.00653616046396089\\
};
\addplot [color=red, forget plot]
  table[row sep=crcr]{%
0	0.76638860859402\\
0.0339185118425725	0.76638860859402\\
};
\addplot [color=red, draw=none, mark=o, mark options={solid, red}, forget plot]
 plot [error bars/.cd, y dir = both, y explicit]
 table[row sep=crcr, y error plus index=2, y error minus index=3]{%
0.003	0.815412025623549	0.0042933672257855	0.0042933672257855\\
0.0368039514216498	0.814511409327325	0.00525341892087424	0.00525341892087424\\
0.017804337790495	0.814334881288606	0.00511516105670108	0.00511516105670108\\
0.0110084341802039	0.818604149372734	0.00657673388943804	0.00657673388943804\\
};
\addplot [color=red, forget plot]
  table[row sep=crcr]{%
0	0.815412025623549\\
0.0338039514216497	0.815412025623549\\
};
\addplot [color=red, draw=none, mark=o, mark options={solid, red}, forget plot]
 plot [error bars/.cd, y dir = both, y explicit]
 table[row sep=crcr, y error plus index=2, y error minus index=3]{%
0.003	0.846604293174005	0.00361952609044099	0.00361952609044099\\
0.0292128115098897	0.844986629859708	0.00378534097514585	0.00378534097514585\\
0.0174128053797795	0.848221459705486	0.0037847596932673	0.0037847596932673\\
};
\addplot [color=red, forget plot]
  table[row sep=crcr]{%
0	0.846604293174004\\
0.0262128115098897	0.846604293174004\\
};
\addplot [color=red, draw=none, mark=o, mark options={solid, red}, forget plot]
 plot [error bars/.cd, y dir = both, y explicit]
 table[row sep=crcr, y error plus index=2, y error minus index=3]{%
0.003	0.868892069421505	0.00317666497967072	0.00317666497967072\\
0.0492592532895846	0.869085205889603	0.00324071874256917	0.00324071874256917\\
0.028808370048028	0.868698122760476	0.00324750891442106	0.00324750891442106\\
};
\addplot [color=red, forget plot]
  table[row sep=crcr]{%
0	0.868892069421505\\
0.0462592532895846	0.868892069421505\\
};
\addplot [only marks,color=red, draw=none, mark=o, mark options={solid, red}]
 plot [error bars/.cd, y dir = both, y explicit]
 table[row sep=crcr, y error plus index=2, y error minus index=3]{%
0.003	0.886984265650111	0.00277410110423164	0.00277410110423164\\
0.048836021665074	0.886984265650111	0.00277410110423184	0.00277410110423184\\
};
\addlegendentry{$\slope=0$}

\addplot [color=red, forget plot]
  table[row sep=crcr]{%
0	0.886984265650111\\
0.045836021665074	0.886984265650111\\
};
\addplot [color=red, dashed, forget plot]
  table[row sep=crcr]{%
0.05	0.75\\
0.05	0.9\\
};
\end{axis}
\end{tikzpicture}%
%
\caption{Continuum limit \cref{eq:continuumfit} of the step scaling function with ($\slope\ne0$) and without ($\slope=0$) slope. Red markers are shifted for visualization.}
\label{fig:continuumfit}
\end{wrapfigure}

In an iterative process one computes the step scaling function 
\begin{equation}
  \gqqt(sr)=\sigma(s,\gqqt(r))\,,\quad s=0.75
\end{equation}
with scale factor $s$. The step scaling function $\sigma$ is a discrete $\beta$ function. Starting at a given point $(\gqqt(r_*)=u_0)$ a series 
is formed by applying the step scaling function iteratively:
\begin{align*}
  u_0  & =\gqqt(r_*) \qquad r_*=0.45r_0\\
  u_1  & =\gqqt(sr_*)=\sigma(s,u_0) \\
  u_2  & =\gqqt(s^2r_*)=\sigma(s,u_1)\\
    & \qquad\vdots\\
  u_5&=\gqqt(s^{5}r_*)=\sigma(s,u_4)
\end{align*}
In this way one can reach down from $r_*=0.45r_0$ to $r_5=s^5r_*\approx0.11r_0$, visualized in \cref{fig:step_scaling_iter}.
In each iteration one has to compute the lattice equivalent $\Sigma$ of the step scaling function, which has an additional dependence on the lattice spacing $a$,
\begin{align}   
  \gqqt(sr,a)&=\Sigma(s,u,a/r) \left. \right|_{\gqqt(r,a)=u}
\end{align}
and perform its continuum extrapolation, 
		\begin{align}
		\sigma(s,u)=\lim_{a/r\to0}\Sigma(s,u,a/r)\,,
		 \end{align}
		 which is the starting point of the next iteration. The extrapolation to $a/r\to0$ is performed as a linear fit
		 \begin{align}
		 	\Sigma(s,u,a/r)=\sigma(s,u)\{ 1+\rho\left(a/r\right)^2\}
		 	\label{eq:continuumfit}
		\end{align}
		with slope $\rho(u)$ and continuum value $\sigma(s,u)$.
		To test our treatment of cut off effects we extrapolate	with and without slope $\rho$, where the extrapolation without slope is constrained to data points $(a/r)^2\le0.05$  
		close to the continuum. The two different extrapolations can be seen in \cref{fig:continuumfit}.  The five different pairs of black solid and red dotted lines in \cref{fig:continuumfit} show the fit of $\Sigma(s,u,a/r)$ scaled by $1/u$. The uppermost pair corresponds to the continuum extrapolation of the last iteration. Comparing the results for $\rho=0$ and $\rho\ne0$, indicates that cut off effects are under good control. We use the ones with larger errors 
		($\rho\ne0$) for further analysis.

		In the small distance regime the step scaling strategy is beneficial in comparison to the traditional continuum limit, in which one would have to compute the coupling up to $r\ge r_0$. With step scaling the essential measurements on fine lattices involve only short distance quantities, where self-averaging 
works very well. This reduces computational requirements, less statistics is needed.

		The continuum extrapolated non-perturbative values of $\sigma(s,u)$ can be compared to the prediction of perturbation theory. The latter are obtained by inserting
		the perturbative $\beta$-function into
		\begin{equation}
		 	\ln (s)=-\int\limits_{\sqrt{u}}^{\sqrt{\sigma(s,u)}}\frac{1}{\beta^{\rm qq}(g)}\operatorname{d}g
		 \end{equation} 
		 and solving for $\sigma(s,u)$. 
		 The comparison in \cref{fig:stepfun_comp} shows surprisingly clear deviations from perturbation theory. 
		 The non-perturbative $\sigma(s,u)$ crosses the 4-loop
		 prediction at around $u=3.5$ and is significantly lower
		 for $u=2.4\,$.

\section{The $\Lambda$-parameter} 
\label{sec:the_lambda}
The $\Lambda$ parameter was calculated from lattice data 
\begin{figure}[t]
	\centering
	\setlength\figureheight{0.25\linewidth} 
	\setlength\figurewidth{0.3\linewidth}
  \begin{subfigure}{0.48\textwidth}
  \input{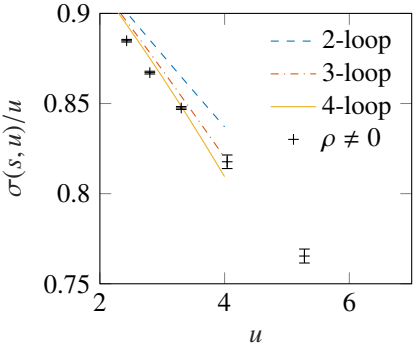}
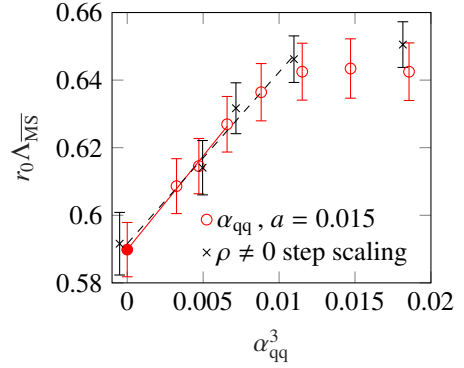
      \caption{The non-perturbative  $\sigma(s,u)$ compared with its perturbative equivalent up to four loop order.}
      \label{fig:stepfun_comp}
  \end{subfigure}\hfill
  \begin{subfigure}{0.48\textwidth}
%
\begin{tikzpicture}

\begin{axis}[%
width=0.988\figurewidth,
height=\figureheight,
at={(0\figurewidth,0\figureheight)},
scale only axis,
xmin=-0.001,
xmax=0.02,
xlabel style={font=\color{white!15!black}},
xlabel={$\aqq^3$},
ymin=0.58,
ymax=0.66,
ylabel style={font=\color{white!15!black}},
ylabel={$r_0\LMS$},
axis background/.style={fill=white},
legend style={at={(0.97,0.03)}, anchor=south east, legend cell align=left, align=left, fill=none, draw=none},
scaled ticks=false,xtick={0, 0.005, 0.01, 0.015, 0.02},xticklabels={0, 0.005, 0.01, 0.015, 0.02},tick label style={/pgf/number format/fixed}
]
\addplot [only marks,color=red, draw=none, mark=o, mark options={solid, red}]
 plot [error bars/.cd, y dir = both, y explicit]
 table[row sep=crcr, y error plus index=2, y error minus index=3]{%
0.00324574430774656	0.608626035312787	0.00812246350386618	0.00812246350386618\\
0.00471767175011834	0.614553120650076	0.00817537312537026	0.00817537312537026\\
0.00657641335278576	0.626936586744235	0.00821969279804958	0.00821969279804958\\
0.00882071147734426	0.636423160775302	0.00847448272807946	0.00847448272807946\\
0.0115231312679017	0.642495389649845	0.0084215625561939	0.0084215625561939\\
0.0147056882140278	0.643434608325699	0.00877998820585682	0.00877998820585682\\
0.0185575906379285	0.642489529298099	0.00852155712138243	0.00852155712138243\\
0.0233629543210293	0.641305988763482	0.00952660107873992	0.00952660107873992\\
0.0290320595932299	0.636063056614145	0.0138196087292493	0.0138196087292493\\
0.0357083963829069	0.627823515476893	0.0108063333095384	0.0108063333095384\\
0.0475316889568334	0.634306253865085	0.00932655216686658	0.00932655216686658\\
0.0570262869594281	0.618319062678196	0.00606430274888424	0.00606430274888424\\
0.0728317190070102	0.611446076631167	0.0104353447376029	0.0104353447376029\\
};
\addlegendentry{$\aqq\,$, $a=0.015$}

\addplot [color=red, forget plot]
  table[row sep=crcr]{%
0	0.589843877372089\\
0.00657641335278576	0.626293898231212\\
};
\addplot [color=red, mark=*, mark options={solid, fill=red, red}, forget plot]
 plot [error bars/.cd, y dir = both, y explicit]
 table[row sep=crcr, y error plus index=2, y error minus index=3]{%
0	0.589843877372089	0.00802989672356009	0.00802989672356009\\
};
\addplot [only marks,color=black, draw=none, mark=x, mark options={solid, black}]
 plot [error bars/.cd, y dir = both, y explicit]
 table[row sep=crcr, y error plus index=2, y error minus index=3]{%
0.00496312211221777	0.614079863060214	0.00804161618888673	0.00804161618888673\\
0.00715880624373053	0.63168012412576	0.00752670783766896	0.00752670783766896\\
0.0109793116155257	0.646196857171363	0.00688744886898135	0.00688744886898135\\
0.0181524720657869	0.650510544046621	0.00676113637125249	0.00676113637125249\\
0.0332175396090857	0.637534266060737	0.00547423322133353	0.00547423322133353\\
0.0741828815987871	0.606306505462578	0.00304338202139832	0.00304338202139832\\
};
\addlegendentry{$\rho\ne0$ step scaling}

\addplot [color=black, dashed, forget plot]
  table[row sep=crcr]{%
0	0.591589358018519\\
0.0109793116155257	0.647339799447924\\
};
\addplot [color=black, mark=x, mark options={solid, fill=black, black}, forget plot]
 plot [error bars/.cd, y dir = both, y explicit]
 table[row sep=crcr, y error plus index=2, y error minus index=3]{%
-0.0005	0.591589358018519	0.00926182352908375	0.00926182352908375\\
};
\end{axis}
\end{tikzpicture}%
  \vspace*{-0.3cm}
\caption{Extrapolation of the $\Lambda$ parameter from the 4-loop $\beta$-function in the \qq-scheme.}
\label{fig:extrapolateLMS}
  \end{subfigure}
  \caption{The running of the coupling and the $\Lambda$-parameter.}
\end{figure}
\begin{equation}
  \Lqq=\varphi(\gqq(r))/r\,,\qquad\varphi(\gqq(r))=(b_{0} \gqq ^{2}) ^{-\frac{ b_{1}}{2 b_{0}^{2}}}
        \mathrm{e}^{-\frac{1}{2 b_{0}\gqq ^{2}}} \times
        \exp{
        \left[
          -\int\limits_{0}^{\gqq}\,{\rm d}{x}
          \left(
            \frac{1}{ \beta^{\rm qq}(x)}+\frac{1}{ b_{0}x^{3}}- \frac{ b_{1}}{ b_{0}^{2}x}
          \right)
          \right]}
          \label{eq:lambda_parameter}
\end{equation}
using the perturbative $\beta$-function at 4-loop order. As a Renormalization Group Invariant the $\Lambda$-parameter should not change as we vary $\gqqt(r)$. Of course this only holds true in a regime where the 4-loop
approximation of $\beta^{(\qq)}$ is sufficiently accurate, i.e. the coupling is sufficiently small. 
We have apparently not reached this region, but
by extrapolating the last three values of $r_0\LMS$ to $\gqq\to0$ it is possible to determine the non-perturbative value of $\LMS$. \Cref{fig:extrapolateLMS} shows the extrapolation
\begin{equation}
  r_0\Lqq= r_0\lim_{\substack{r\to0\\(\gqq\to 0)}}\varphi(\gqq(r))/r
\end{equation}
for the coupling $\aqq(r)=\gqqt(r)/4\pi$ in the continuum and at finest lattice spacing.

The last three points are used to extrapolate linearly to $\aqq=0$. The final result $r_0\LMS=0.590(16)$ is from the $\rho\ne0$ extrapolation shown in \cref{fig:extrapolateLMS}, taking also the 
results with $\rho=0$ into account to estimate
the final range. 

At $\aqq=0.24$ ($\aqq^3\approx0.014$), the result $r_0 \LMS^\text{4-loop}$ deviates significantly from the extrapolated value. A similar observations was made in \cite{PhysRevLett.117.182001} at $\alpha=0.19$ 
in a different non-perturbative renormalization scheme.

\section{Conclusion} 
\label{sec:conclusion}

Open boundary conditions prevent topological freezing when simulating fine lattice spacings. This allows the measurement of short distance and small flow-time quantities up to a regime, in which one can test perturbation theory and the small flow-time expansion. We computed the coupling in the \qq-scheme down to about
$\alpha_\mathrm{qq}=0.2$. Extracting the $\Lambda$-parameter 
from these coupling values using the four loop $\beta$-function
still showed a significant dependence on $\alpha$. 
Linear extrapolation in the (perturbatively) dominating correction
term $\alpha^3$ yielded our estimate 
\begin{align}   
 r_0\LMS=0.590(16)\,.
\end{align}
It is in agreement with the FLAG average \cite{Aoki:2013ldr}
but has a smaller error. The extrapolation to $\alpha=0$ in 
figure~\ref{fig:extrapolateLMS} pushes our estimate
some 1-2 standard deviations 
below previous pure gauge theory results \cite{Brambilla:2010pp,Gockeler:2005rv} with small errors.

Our investigation of the behaviour of $\Exp{E(t,x_0)}$ 
near the open boundary showed that effects 
are noticeable up to a distance of about $x_0=\sqrt{20t_0}$ ($\approx 0.8$fm), roughly independent of the lattice spacing.
Hence boundary improvement will not appreciably enlarge the usable fraction 
of lattices with open boundary conditions.

As a proof of concept we studied the 
small flow-time limit of 
the action density near the boundary. As $t\to 0$
it yields a matrix element of the trace of the energy momentum tensor. The expected linear behaviour in $t$ is indeed found and the zero flow-time limit seems feasible. Hence the method can be applied, e.g. for finite temperature physics \cite{Ejiri_LatentHeat,Suzuki_EnergyMomentumTensor}.
Due to the distinct discretisation effects at small flow time 
as well as a significant $t$-dependence, we do find
that continuum extrapolation and $t\to0$ extrapolation 
are necessary -- of course in the correct order.
Note that for $\alpha_\mathrm{qq}$ we had lattice spacings
as small as $a=0.015\,\fm$, while for the small flow-time 
expansion we only went down to $a=0.025\,\fm$ so far.

\clearpage
\bibliography{lattice2017}

\end{document}